\documentclass[11pt,twoside]{article}

\usepackage{asp2006}
\usepackage{epsf}

\begin{document}
\title{Mid-Infrared Spectra of Radio Galaxies and Quasars}   
\author{P. M. Ogle}   
\affil{Spitzer Science Center, Caltech, 220-6, Pasadena, CA 91125}    
\author{R. R. J. Antonucci}
\affil{Physics Department, University of California, Santa Barbara, CA 93106}
\author{D. Whysong}
\affil{NRAO, Array Operations Center, P.O. Box O, 1003 Lopezville Road, Socorro, NM 87801-0387}

\begin{abstract} 
Spitzer Infrared Spectrograph (IRS) observations of 3C radio galaxies and quasars shed new light 
on the nature of the central engines of AGN. Emission from silicate dust obscuring the central 
engine can be used to estimate the bolometric luminosity of an AGN. Emission lines from ions such 
as O {\sc iv} and Ne {\sc v} give another indication of the presence or lack of a hidden 
source of far-UV photons in the nucleus. Radio-loud AGN with relative-to-Eddington luminosity 
ratios of $L/L_\mathrm{Edd}<3\times10^{-3}$ do not appear to have broad optical emission 
lines, though some do have strong silicate emission. Aromatic emission features from star formation 
activity are common in low-luminosity radio galaxies. Strong H$_2$ pure-rotational emission lines
are also seen in some mid-IR weak radio galaxies, caused by either merger shocks or jet shocks in 
the interstellar medium. 

\end{abstract}



\section{Mid-IR Spectra}

We report on {\it Spitzer} IRS spectrophotometry of 72 3C radio galaxies and quasars. 
Our survey of a complete (redshift $z<1.0$ and  $S_\nu(178~\mathrm{MHz}) > 16.4$ Jy) sample of 52 Fanaroff 
and Riley (FR) type II (edge-brightened) radio galaxies and quasars has been published \citep*{owa06}.  
We are conducting a complementary survey of 20 lower luminosity ($z<0.1$) FR I radio galaxies  to determine 
the nature of their central engines. Low resolution spectra of FR I radio galaxies were taken with an exposure 
time of 480-1440 s per spectral order. Radio galaxies show a wide variety of mid-IR spectral features and can 
be classified by the strength of those features  (Fig. 1). 

Broad-line radio galaxies (and quasars) such as 3C 120 are characterized by high-ionization emission lines, 
including [O {\sc iv}], [Ne {\sc v}], [S {\sc iv}] and [Ne {\sc vi}], and strong silicate dust emission 
bumps at 10 and 18 $\mu$m. The high-ionization lines are powered by far-UV photons from a
luminous accretion disk. The silicate emission is produced by dust out of the line of sight which absorbs
a large fraction ($\sim 20\%$) of the luminosity from the accretion disk. Though 3C 120 has
a relativistic radio jet viewed at a small angle ($< 20\deg$), the mid-IR spectrum does not appear 
to contain a large fraction of synchrotron emission. In contrast, the spectrum of BL Lac is a pure power-law, 
presumably dominated by synchrotron emission from its relativistically beamed jet.

Mid-IR weak radio galaxies such as NGC 6251 and 3C 270 have low-ionization emission line spectra
from [Ne {\sc iii}], [Ne {\sc ii}], and [Ar {\sc ii}]. There is no strong source of UV photons from the nuclei
to power higher ionization lines such as [Ne {\sc v}] (at least at a detectable level). Curiously, however,
there are strong silicate emission features that dominate the mid-IR continuum. Perhaps the AGN continuum
is powerful enough to warm circumnuclear dust, even though it is not powerful enough to produce high-ionization
lines or a broad line region.

In some sources, such as 3C 31 and 3C 293, large equivalent-width aromatic features from polycyclic aromatic
hydrocarbons (PAHs) indicate a large starburst fraction. Any AGN contributions to the spectra are difficult to 
measure because they are significantly weaker than the starbursts. PAH features (especially at 11.3 $\mu$m) are 
also visible at lower equivalent width in most other radio galaxies (except BL Lac). Evidently, the radio 
galaxy hosts are not dead ellipticals but instead contain a rich ISM capable of forming new stars. 

\begin{figure}[!ht]
\plotone{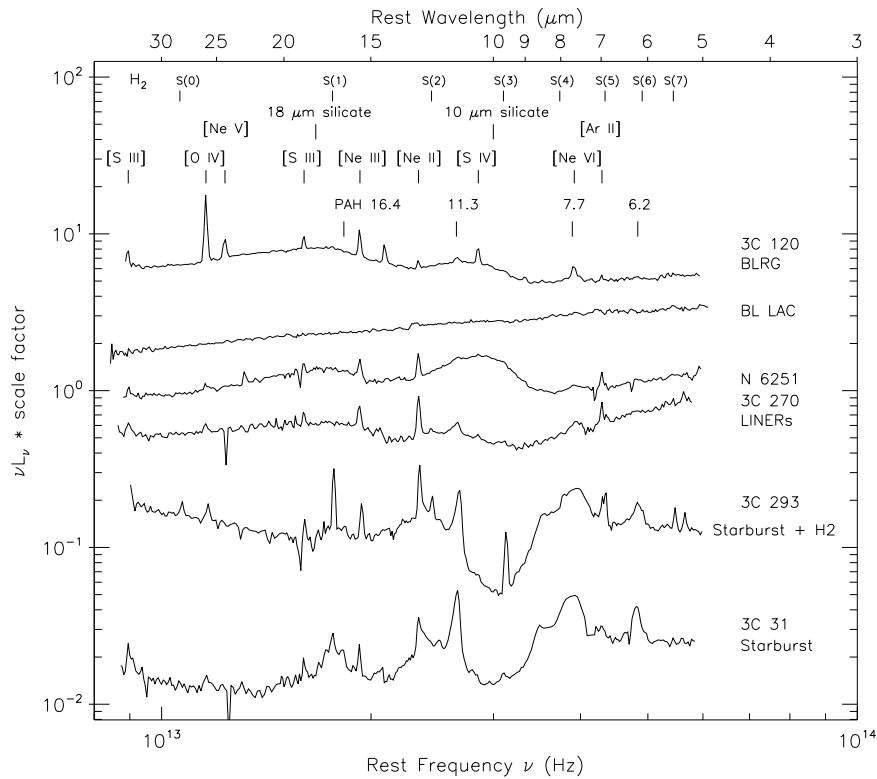}
\caption{{\it Spitzer} IRS spectra of radio galaxies. Sources range from
         AGN dominated at the top to starburst dominated at the bottom. 
         High ionization lines are excited by far-UV continuum from the 
         accretion disk in 3C 120. BL Lac has a power-law synchrotron emission spectrum.
         LINER galaxies NGC 6251 and 3C 270 have strong silicate emission
         features at 10 and 18 $\mu$m. Very strong H$_2$ lines are seen from 
         3C 293. Polycyclic aromatic hydrocarbon features from star formation
         activity dominate the spectrum of 3C 31.}
\end{figure}

\section{AGN Luminosities}

Mid-IR spectroscopy is a powerful way to estimate the bolometric luminosities
of AGNs, unhampered by extinction effects \citep{wa04, owa06}. The ratio of mid/far-IR 
emission from 3C quasars and radio galaxies varies by less than a factor of
3 \citep{srh05}, indicating very little mid-IR absorption or anisotropy.
However, a plot of $\nu L_\nu (15~\mu\mathrm{m})$ vs. $\nu L_\nu (178~\mathrm{MHz})$ shows
a range of 3 orders of magnitude in mid-IR luminosity for a given radio luminosity (Fig. 2).
Radio jet power is not tightly correlated with accretion disk luminosity, so some other 
parameter such as black hole spin may be important for driving radio jets.

Considering first the FR II narrow-line radio galaxies (NLRGs) in Fig. 2, we see that 
only 50\% have mid-IR luminosities comparable to quasars and broad-line radio galaxies 
($\nu L_\nu (15\mu\mathrm{m})>10^{44}$ erg s$^{-1}$). The rest are mid-IR weak,
even though they have powerful radio jets \citep{owa06}. Mid-IR weak FR IIs are a factor of $\sim$ 10 
brighter in the mid-IR than the average FR I radio galaxy, and may be considered higher-luminosity
cousins to FR Is. 

Most of the FR I radio galaxies in our sample (except 3C 120 and Per A) are mid-IR weak and
have no broad emission lines (Fig. 2).  Several authors have suggested that FR I jets are powered by extremely 
sub-Eddington, radiatively inefficient accretion flows \citep{rdf96}. To test this hypothesis, we estimate the 
mass of the central supermassive black hole from the host K-band stellar luminosity and use this together with
the IR luminosity to estimate the Eddington ratio. We find that all mid-IR weak FR Is (and FR IIs) have 
$L/L_\mathrm{Edd}<3\times10^{-3}$, and no broad emission lines in their optical spectra, 
consistent with the absence of a luminous, optically thin accretion disk.


\begin{figure}[!ht]
\plotfiddle{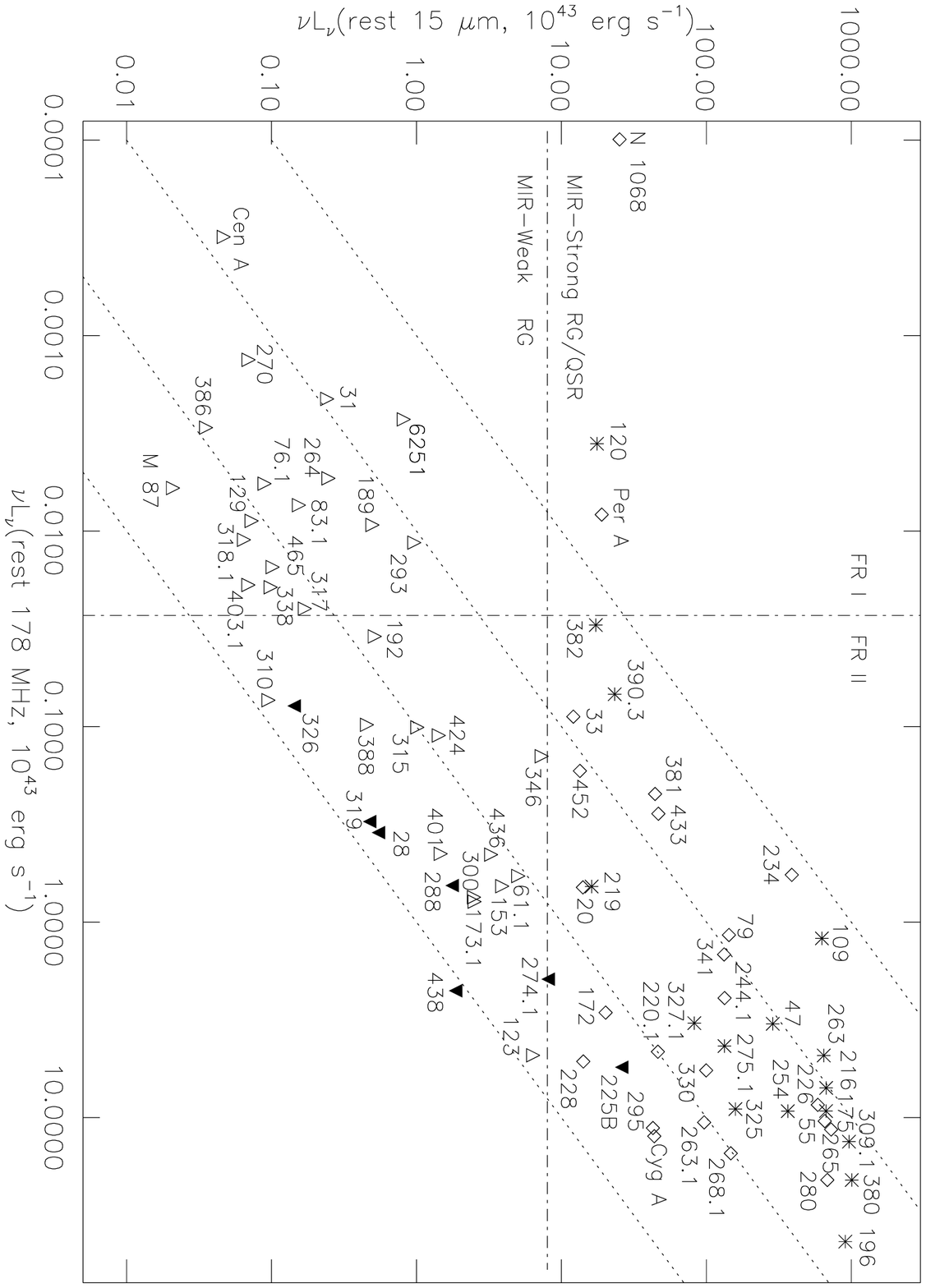}{3.2in}{90}{50}{50}{180}{-30}
\caption{Mid-IR luminosity at 15 $\mu$m vs. radio luminosity at 178 MHz.
         Quasars (asterisks) and obscured type 2 quasars (diamonds) have
         $\nu L_\nu (15\mu\mathrm{m})>10^{44}$ erg s$^{-1}$. Mid-IR weak
         FR IIs and FR Is (triangles, filled $=$ upper limit) have lower 
         luminosities and none of them have broad emission lines. For
         comparison, the Seyfert 2 galaxy NGC 1068 is also shown.}  
\end{figure}

\section{H$_2$ Emission}

Roughly 20\% of the mid-IR weak radio galaxies have very large equivalent width H$_2$ emission lines
in their spectra (e.g. 3C 293, Fig. 1). In contrast, very few mid-IR luminous radio galaxies have detectable 
H$_2$ emission (e.g. Per A and 3C 433), and the lines have low equivalent width. We suggest that the H$_2$ 
emission is therefore {\it not} powered by X-rays from the AGN. 

The galaxy 3C 293 is the brightest example of a radio galaxy with strong H$_2$ emission. An excitation diagram 
of the H$_2$ pure rotational levels indicates a range of gas temperature from 300-1400 K. The luminosity of 
the mid-IR H$_2$ lines sums to $6.0\times10^{41}$ erg s$^{-1}$, from a warm molecular hydrogen mass of 
$3\times10^{8}$ M$_\odot$. Radio observations of CO emission indicate roughly $2\times 10^{10}$ M$_\odot$ of 
cold molecular hydrogen in 3C 293 \citep{ess99}. The host galaxy is very dusty and has a distorted morphology 
indicating a recent interaction with another galaxy. In addition, high velocity H {\sc i} absorption and 
N {\sc ii} emission \citep{moe03,emt05} indicate a strong interaction between the radio jet and host galaxy ISM.

Because of the rich phenomenology of H$_2$ emitting radio galaxies, it has proven difficult to determine the 
primary power source. It is likely that H$_2$ is thermally excited in slow (non-radiative) shocks in the ISM. 
These shocks could be from molecular gas accretion from companions or the IGM, or jet shocks. We are pursuing 
narrow-band imaging and spectroscopy of near-IR H$_2$ ro-vibrational lines to isolate the power source.

\acknowledgements 
This work is based on observations made with the {\it Spitzer} Space Telescope, 
operated by the Jet Propulsion Laboratory, California Institute of Technology under NASA 
contract 1407. Support for this research was provided by NASA through an award issued by JPL/Caltech. 



\begin{thebibliography}{}
\bibitem[Emonts et al. (2005)]{emt05} Emonts, B. H. C., Morganti, R., Tadhunter, C. N., Osterloo, T. A., Holt, J., 
\& van der Hulst, J. M. 2005, MNRAS, 362, 931
\bibitem[Evans et al. (1999)]{ess99} Evans, A. S., Sanders, D. B., Surace, J. A., \& Mazzarella, J. M. 1999, ApJ, 511, 730.
\bibitem[Morganti et al. (2003)]{moe03} Morganti, R., Osterloo, T. A., Emonts, B. H. C., van der Hulst, J. M., \& Tadhunter, C. N. 2003, ApJ, 593, L69. 
\bibitem[Ogle et al. (2006)Ogle, Whysong, \& Antonucci]{owa06} Ogle, P., Whysong, D., \& Antonucci, R. 2006, ApJ, 647, 161
\bibitem[Shi et al. (2005)]{srh05} Shi, Y. et al. 2005, ApJ, 629, 88
\bibitem[Whysong \& Antonucci (2004)]{wa04} Whysong, D. H., \& Antonucci,
R. R. J. 2004, ApJ, 602, 116
\bibitem[Reynolds et al. (1996)]{rdf96} Reynolds, C. S., Di Matteo, T., Fabian, A. C., Hwang, U., \& Canizares, C. R.
1996, MNRAS, 283, L11

\end{thebibliography}
\end{document}